\input harvmac
\newcount\figno
\figno=0
\def\fig#1#2#3{
\par\begingroup\parindent=0pt\leftskip=1cm\rightskip=1cm
\parindent=0pt
\baselineskip=11pt
\global\advance\figno by 1
\midinsert
\epsfxsize=#3
\centerline{\epsfbox{#2}}
\vskip 12pt
{\bf Fig. \the\figno:} #1\par
\endinsert\endgroup\par
}
\def\figlabel#1{\xdef#1{\the\figno}}
\def\encadremath#1{\vbox{\hrule\hbox{\vrule\kern8pt\vbox{\kern8pt
\hbox{$\displaystyle #1$}\kern8pt}
\kern8pt\vrule}\hrule}}

\overfullrule=0pt

\Title{{\it TIFR-TH/97-05}}
{\vbox{\centerline{The Effectiveness of D-branes in the}
\centerline{Description
of Near-Extremal Black Holes}}}
\smallskip
\centerline{Sumit R. Das\foot{E-mail: das@theory.tifr.res.in}}
\smallskip
\centerline{\it Tata Institute of Fundamental Research}
\centerline{\it Homi Bhabha Road, Bombay 400 005, INDIA}
\smallskip
\bigskip

\medskip

\noindent

It is known that weak coupling calculations of absorption or emission
by slightly non-extremal D-brane configurations are in exact agreement
with semiclassical results for the black holes they describe at strong
couplings.  We investigate one open string loop corrections to
processes involving single and parallel
D-branes and show that a class of relevant
terms vanish, indicating that these processes are not
renormalized. Our results have implications for five dimensional black
holes and extremal 3-branes.
 
\Date{March, 1997}

\def\ap{\alpha'}
\def\half{{1\over 2}}
\def\Te{T_{eff}}

\bigskip
\bigskip

\newsec{Introduction}

Recently the idea that massive string states become black holes
when the coupling is large \ref\sussk{L. Susskind, hep-th/9309145;
J. Russo and L. Susskind, Nucl. Phys. B 437 (1995) 611}
\ref\duff{ M. Duff and
J. Rahmfeld, Phys. Lett. B345 (1995) 441, hep-th/9406105}
\ref\sen{ A. Sen,
Nucl. Phys. B440 (1995) 421 , hep-th/9411187 and Mod. Phys. Lett.
A10 (1995) 2081.} has been very successful.
In particular, five dimensional extremal black holes with large horizons are
described by bound states of D-branes whose degeneracies exactly
reproduce the Beckenstein-Hawking entropy 
\ref\stromvafa{A. Strominger and C. Vafa, {\it Phys.
Lett.} B379 (1996) 99,
hep-th/9601029.}. This result was extended to four dimensional
extremal black holes \ref\fdh{J. Maldacena and A. Strominger,
Phys. Rev. Lett. 77 (1996) 428, hep-th/9603060; C. Johnson, R. Khuri
and R. Myers, Phys. Lett. B378 (1996) 78, hep-th/9603061} and to
spinning black holes \ref\spin{J. Breckenridge, R. Myers, A. Peet and
C. Vafa, hep-th/9602065; J. Breckenridge, D. Lowe, R. Myers, A. Peet,
A. Strominger and C. Vafa, Phys. Lett. B381 (1996) 423}.

The entropy and hawking temperature continue to agree in the near
extremal limit
\ref\callanmal{C. Callan and J. Maldacena, Nucl. Phys. B 475 (1996) 645,
hep-th/9602043.},
\ref\horostrom{G. Horowitz and A. Strominger, Phys. Rev. Lett.
77 (1996) 2368, hep-th/9602051.}. 
It was also found that the lowest order decay rate of slightly nonextremal 
D-brane configurations is proportional to
the horizon area \callanmal\ which is consistent with the semiclassical
Hawking radiation from such holes, as shown in 
\ref\dmw{A. Dhar, G. Mandal and S.R. Wadia, 
Phys. Lett. B388 (1996) 51, hep-th/9605234.}. 
Rather surprisingly the two emission (and
absorption) rates for neutral scalars 
were in fact found to agree {\it exactly} 
\ref\dmatb{S.R. Das and S.D. Mathur, Nucl. Phys. B478 (1996) 561,
hep-th/9606185; Nucl. Phys. B482 (1996) 153, hep-th/9607149.}. Exact
agreements were also found for neutral and charged scalars in five
and four dimensional black holes \ref\gubkle{S. Gubser and
I. Klebanov, Nucl. Phys. B482 (1996) 173, hep-th/9608108.}.
Even more remarkably, the grey body factors which describe a nontrivial energy
dependence of the absorption cross-section 
at higher energies are also in exact agreement
\ref\malstrom{J. Maldacena and A. Strominger, Phys. Rev. D55 (1997) 861,
hep-th/9609026.}, a result verified in the four dimensional case as
well \ref\gubkleb{S. Gubser and I. Klebanov, Phys. Rev. Lett. 77
(1996) 4491, hep-th/9609076.}. A general analysis of
classical
absorption by such black holes and the possibility of agreement of 
classical and D-brane greybody factors, has been carried out in
\ref\dealwis{S.P. de Alwis and K.Sato, hep-th/9611189}.

A more detailed test of these ideas is provided by the emission and
absorption of ``fixed'' scalars by five dimensional black holes
\ref\fixed{S. Ferrara, R. Kallosh and A. Strominger,
Phys. Rev. D52 (1995) 5412, hep-th/9508072.;
S. Ferrara and R. Kallosh, hep-th/9602136;
hep-th/9603090;  G. Gibbons, R. Kallosh
and B. Kol, hep-th/9607108; B. Kol and A. Rajaraman, hep-th/9608126.} 
, where agreement between
the D-brane and classical calculations were demonstrated in 
\ref\callanfixed{C.Callan, S. Gubser, I. Klebanov and
A. Tseytlin, hep-th/9610172.}. The grey body factors at higher energies
agree as well 
\ref\klebkras{I. Klebanov and M. Krasnitz, hep-th/9612051}.

A related example where there seem to be exact agreement of D-brane
and general relativity results is the absorption of $l =0$ and $l = 1$
waves by 
extremal 3-branes with no momentum\ref\igorc{I. Klebanov, hep-th/9702076}
\ref\igore{S. Gubser, 
I. Klebanov and A. Tseytlin, hep-th/9703040}. The cross-sections for
higher partial waves agree upto numerical factors \igore. In this case
the near-extremal entropy also differs by a numerical factor \ref\igord
{S. Gubser, I. Klebanov and A. Peet, Phys. Rev. D54 (1996) 3915,
hep-th/9602135; I. Klebanov and A. Tseytlin, Nucl. Phys. B475 (1996)
165, hep-th/9604089}.

For systems arbitrarily far from extremality (like the Schwarzschild
black hole) it has been argued that the microscopic and semiclassical
answers should match only at a special value of the coupling at which
the horizon curvature is of the string scale \sussk,
\ref\horpol{G. Horowitz and J. Polchinski, hep-th/9612146.}. It was
shown in \horpol\ that in all known cases the stringy and
semiclassical entropies indeed match at this point, upto numerical
factors. 

The reason why these results appear surprising is that the D-brane
calculations are performed in the lowest order of the open string
perturbation theory, using in fact a low energy effective action,
while the brane configurations describe large black holes only
when the open string coupling is large. For extremal BPS holes, 
nonrenormalization theorems allow an extrapolation of weak coupling
results to strong coupling. However away from extremality there does
not appear to be an obvious reason why this can be done.

Consider for example the five dimensional near
extremal black hole which is described by bound states of 1D-branes and
5D-branes with some momentum flowing along the 1D brane. Non extremality
is introduced by allowing both left and right momenta on the 1D brane
and the classical solution of the low energy effective action has 
a ten-dimensional string metric
\eqn\bone{\eqalign{
ds^2 & = (1+{r_1^2 \over r^2})^{-1/2}(1+{r_5^2 \over r^2})^{-1/2}[
dt^2 - dx_5^2 -{r_0^2 \over r^2}({\rm cosh}\sigma~dt + {\rm
sinh}\sigma
~dx_5)^2]\cr
&- (1+{r_1^2 \over r^2})^{1/2}(1+{r_5^2 \over r^2})^{-1/2}
[dx_1^2 + dx_2^2 + dx_3^2 + dx_4^2] \cr
&- (1+{r_1^2 \over r^2})^{1/2}(1+{r_5^2 \over r^2})^{1/2}
[(1 - {r_0^2 \over r^2})^{-1} dr^2 + r^2 d\Omega_3^2]}}
The various length scales are given in terms of the charges by
\eqn\btwo{\eqalign{
& r_1^2  =  {16 \pi^4 \ap^3 (g Q_1) \over V}~~~~~~~~~~~~~~~~~~
 r_5^2  = \ap (gQ_5)\cr
& \half r_0^2 {\rm sinh} 2\sigma  =  { 16 \pi^4 \ap^4 (g^2 N) \over R^2 V}
~~~~~~~~~~~~~~~~ r_N^2  =  r_0^2 {\rm sinh}^2 \sigma}}
Here $\ap = 1/(2\pi T)$ where $T$ is the elementary string tension,
$g$ is the string coupling. The brane configuration lies on a $T^4 \times S^1$
with the one brane along the $S^1$. The radius of this $S^1$ is $R$, while
the volume of the $T^4$ is $V$. The integers $Q_1, Q_5, N$ are the 1-brane
RR charge, 5-brane RR charge and the total momentum.
The extremal limit is $r_0 \rightarrow 0$ and $\sigma \rightarrow \infty$
with $r_N$ held fixed. 

It is clear from the classical solution that
the classical limit of the string theory corresponds to $g \rightarrow 0$
with $ gQ_1, gQ_5, g^2 N$ held fixed \malstrom.  In fact we have
large black holes (compared to string scale) when
$  gQ_1, gQ_5, g^2 N > 1 $ and small holes when 
$  gQ_1, gQ_5, g^2 N < 1$. It is in the latter regime that the D-brane
description is good. 

Another example which we will consider in the following is the
self-dual 3-brane in Type IIB theory. The extremal solution has a zero
horizon area, but is completely {\it nonsingular} and the dilaton is a
constant in the classical solution. The extremal string metric is given by
\eqn\btwob{ds^2 = A^{-1/2}(-dt^2 + dx_1^2 + dx_2^2 + dx_3^2)
+ A^{1/2}(dr^2 + r^2 d\Omega_5^2)}
with
\eqn\btwoc{A(r) = 1 + {4\pi \ap^2 (gN) \over r^4}}
where $N$ is the RR charge. The curvature at the horizon $r = 0$ is
$\sim 1/[\ap{\sqrt{gN}}]$. Thus when $gN$ is large, the curvatures are
small and one may trust the supergravity limit. Note that the
parameters in the classical solution depends on $g$ and $\ap$ only
through the combination $\kappa \sim g \ap^2$ where the ten
dimensional Newton constant is $G_{10} = 8 \pi \kappa^2$. For parallel
branes of one type this happens only for 3-branes.

To appreciate the meaning of these products let us consider a simpler
example in scalar $\phi^3$ field theory where we are interested in describing
classical solutions in the presence of a static source \foot{This
analogy was suggested by A. Jevicki.}
\eqn\bthree{
\nabla^2 \phi + \lambda \phi^2 = \kappa \delta ({\vec x})}
By scaling the field $\phi$ it is easily seen that the solution is
of the form
\eqn\bfour{
\phi (x) = {1 \over \lambda} F (x; \lambda \kappa)}
so that the solution is characterized by the product $\lambda \kappa$.
The classical limit is $\lambda \rightarrow 0$ and a solution which
captures the nonlinearity is characterized by $\lambda \kappa = $
fixed. One may obtain the classical solution by summing over all
tree level Feynman diagrams with arbitrary number of insertions of the
source.

The product $\lambda \kappa$ is in fact an exact analog of the products
$gQ$ in the D-brane context in the dilute gas regime
$r_N << r_1, r_5$, as clear from the classical solution. 
The full classical solution can be obtained
by summing over an infinite number of string diagrams which does not
contain any closed string loop, but contains all terms with closed
strings terminating on an aribtrary number of branes. Each such insertion
carries a factor $gQ$ which has to be held finite. In other words we
have to sum over all open string loops.
Closed string loops
do not contain any factor of the charge $Q$ and are therefore suppressed.
This perturbation expansion is a description of the black hole expanded
around flat space-time with the curvature emerging as a result of
summing over open string loops.

It is important to understand why the tree level results are in exact
agreement with answers which are expected to hold for strong
couplings. The point is that even though the near extremal states
break supersymmetry, for slight non-extremality these are states in a
supersymmetric background. In such situations it is natural to expect
that there are nonrenormalization theorems which protect some tree
level results from acquiring loop corrections. Some of these issues
can be addressed in the worldbrane low energy effective theory. Indeed
such considerations have been used by 
Maldacena \ref\Maldacenab{J. Maldacena, hep-th/9611125}
to argue that
processes which follow from the moduli space approximation are not
renormalized to any order. However, these arguments cannot be used to
justify the success of the calculation of other cross-sections like
the absorption or emission of fixed scalars in the five dimensional
black holes and higher partial waves in the 3-brane.

In the following we explicitly examine one open string loop corrections
to some absorption and emission processes from single and parallel
branes. The restriction to parallel branes comes from the fact that it
is for these that we can perform string theoretic calculations
explicitly. Our results are directly relevant to 3-branes. Moreover
since the 1-brane 5-brane system which describes the five dimensional
black hole reduces to a single long D-string, our results have some
relevance to the five dimensional black hole as well. It must be
remembered, though, that the use of an effective string at the loop
level needs further justification.

\newsec{The D-brane calculation}

The D-brane calculation of emission from slightly non-extremal 
configurations is performed in the lowest order of the open string
perturbation theory and at low energies. Let us identify precisely
the approximations used here.

As shown in \ref\maldasuss{ J. Maldacena and L. Susskind, Nucl. Phys
B475 (1996) 679, hep-th/9604042.}  the 1-brane 5-brane system with
large momentum along the string direction must be regarded as multiply
wound along the lines of
\ref\dmata{S.R. Das and S.D. Mathur, Phys. Lett. B375 (1996) 103,
hep-th/9601152.}, \ref\samir{S.D. Mathur, hep-th/9609053.}
so that we essentially have a single long string
which is wrapped around a large circle of radius $Q_1 Q_5 R$. In the
following we will therefore model this system by such a D-string. The
string tension of this effective string should be then fractional $\Te
= 1/( 2\pi \ap Q_5)$ \maldasuss,
\ref\maldaa{ J. Maldacena, Nucl. Phys. B477 (1996) 168, hep-th/9605016.}. 
Furthermore, the massless modes are now
open string modes whose polarizations are restricted to lie along the
$T^4$ directions orthogonal to the string direction $S^1$, together 
with their fermionic partners.

The D-brane calculations are in fact performed in this effective
D-string theory, using the Dirac-Born-Infeld (DBI) effective action
\eqn\cone{
S = {\Te \over 2} \int d^2\sigma~e^{-\phi} {\sqrt{{\rm det}(G_{\mu\nu}^S 
\partial_\alpha X^\mu \partial^\alpha X^\nu)}}}
where $G_{\mu\nu}^S$ is the bulk string frame metric and $\phi$ is the
ten dimensional dilaton. Choosing the static
gauge $\tau = X^0 \equiv x^0, \sigma = X^5 \equiv x^5$
and keeping only the components $X^I$ where $I = 1, \cdots 4$ are
the four $T^4$ directions and retaining terms upto four derivatives on
the open string fields fields $X^I$ the DBI action becomes
\eqn\ctwo{\eqalign{
S = {\Te \over 2} \int dX^0 dX^5 e^{-\phi}[1 + & \half G_{IJ}\partial_+ X^I
\partial_- X^J \cr
& - {1 \over 8} G_{IJ} G_{KL}(\partial_+X^I \partial_+ X^J)
(\partial_-X^K \partial_- X^L)]}}
where $x^\pm = x^0 \pm x^5$. 
The effective action \ctwo\ can be derived by considering 
open string amplitudes on a disc, as verified explicity for the second
term of \ctwo\ in 
\ref\klebahashi{A. Hashimoto and I. Klebanov, Phys. Lett B381
(1996) 437, hep-th/9604065.}.

Note that the metric $G_{IJ}$ is generally a function of all the
coordinates 
\eqn\ctwoa{G_{IJ}[x^0, x^5, X^I(x^0,x^5), X^m(x^0,x^5)]}
(where $m = 6 \cdots 9$), so that
even the term $\partial X^I \partial X^J$ contains an
arbitrary number of open string fields. 
For emission of S-wave modes, the low energy approximation amounts to
ignoring the dependence of the bulk metric on $X^I$ and $X^m$.
(The dependence on $x^0$ and $x^5$ of course
remain which imposes momentum conservation in the Neumann directions).
This amounts to retaining terms which contain the least number of
momenta with a given number of open string fields. We will therefore
expand
\eqn\cthree{G_{IJ} = \delta_{IJ} + {\sqrt{2}}\kappa h_{IJ} (x^0,x^5)}
The BPS states of the system correspond to a background value of only
the left moving part of the $X^I$'s. Non-BPS states correspond to
addition of pairs of oppositely moving modes
\dmata. In the dilute gas
region, and in the lowest order of the open string coupling, one has
a standard thermodynamics problem of a one (spatial) dimensional
gas of bosons and fermions with some given energy and total momentum.
The various thermodynamic quantities like the entropy and the
temperature may be calculated in this gas using standard methods
\dmatb\ and, as is well known by now, for near-BPS
situations the results are in perfect agreement with the semiclassical
answers obtained from the metric \bone\
\callanmal,\horostrom.

For the system of $N$ parallel 3-branes, the extreme low energy
effective worldvolume action is a four dimensional $N=4$ $U(N)$
Yang-Mills theory \ref\wit{E. Witten, Nucl. Phys. B460 (1996) 541,
hep-th/9511030.} with background fields \ref\threeb{R. Leigh,
Mod. Phys. Lett. A4 (1989) 2767; M. Li, Nucl. Phys. B460 (1996) 351;
M. Douglas, hep-th/9512077; A.A. Tseytlin, Nucl. Phys. B469 (1996)
51.}. Emission of higher angular momentum scalars may
be studied by retaining the dependence of the background fields on the
transverse coordinates, as we will see soon. The situation studied so
far is the extremal state with no momentum.
 
\subsec{Minimally coupled scalars for the 1-5 brane}

First consider the emission of minimally coupled scalars. Examples of
these are transverse traceless components of $h_{IJ}$. In the bulk theory
these modes obey the minimally coupled Klein-Gordon equation in the
five dimensional geometry.
For this emission
we can set the dilaton field to zero and in fact concentrate on
one particular component of $h_{IJ}$. The relevant interaction term
which involves the miminum number of derivatives of open string field 
comes from the second term in \ctwo\
\eqn\cfour{
L_{int}^{(1)} = {\sqrt{2}}\kappa h_{IJ} \partial_+ X^I \partial_- X^J}
and describes the annahilation of two oppositely moving open strings 
into a closed string state described by $h_{IJ}$. At low energies,
$h_{IJ}$ is considered as a function of only the longitudinal coordinates 
$x^0, x^5$.

When the outgoing particle does not have any momentum in the $X^5$
direction, it is a neutral scalar from the five dimensional point of
view. The tree level emission rate into a closed string state of
momenta $k$ (note all the $k^i$ with $i = 1 \cdots 5$ are zero), when
averaged over the initial states is \dmatb\ (the notation
is of \malstrom) 
\eqn\cfive{
\Gamma (k) = 2\pi^2 r_1^2 r_5^2 {\pi k_0 \over 2}
{ 1 \over (e^{k_0 \over 2T_L} - 1)(e^{k_0 \over 2T_R}-1)} {d^4 k 
\over (2\pi)^4}}
$T_L$ and $T_R$ are the temperatures of the left and right moving
modes on the effective string. The temperature, which is the same
as the Hawking temperature is then given by 
\eqn\csix{{1 \over T} = {1\over 2}({1 \over T_L} + {1 \over T_R})}
Detailed balance then gives an absorption cross-section
\eqn\cseven{\sigma (k_0) = 2\pi^2 r_1^2 r_5^2 {\pi k_0 \over 2}
{ (e^{k_0 \over T} - 1)
 \over (e^{k_0 \over 2T_L} - 1)(e^{k_0 \over 2T_R}-1)}}
When we restrict to extremely low energies $k_0 << T_L, T_R$ 
(but $k_0 \sim T$) one has $\sigma = A_H$, the area of the horizon,
and this is exactly what we get from a classical absorption
calculation for {\it any} spherically symmetric black hole in {\it
any} number of dimensions 
\ref\dgm{S.R. Das, G. Gibbons and S.D. Mathur, Phys. Rev. Lett.
78 (1997) 417, hep-th/9609052.}.
 
In fact, the
entire expression 
\cseven\ (with $k_0 \sim T_L, T_R$) agrees exactly  
with the classical grey body factor calculated in the dilute gas
regime 
\eqn\bfive{r_0, r_N << r_1, r_5}
and $k_0 \omega_5 << 1$, i.e. at wavelengths much larger than
the gravitational radius \malstrom. 
In fact, beyond the dilute gas approximation detailed
agreements of D-brane and classical grey body factors are not always
present \ref\klmat{I. Klebanov and S.D. Mathur, hep-th/9701187}
\ref\traschen{F. Dowker, D. Kastor and J. Traschen, hep-th/9702109.}.
Even in the dilute gas regime there is some disagreement for charged particle
emission when the individual energies and charges are large, but the
difference is small \traschen.

One important feature of this calculation must be emphasized. Since
the interaction term \cfour\ involves only two open string
fields and the effective tension $\Te$ sits outside the action, the
amplitude is {\it independent of $\Te$}. The decay rate does depend
on the length of the effective string and the agreement with the
semiclassical answer holds when $L_{eff} = 2\pi Q_1 Q_5 R$.

\subsec{Fixed scalars in the 1-brane 5-brane}

A more detailed test is provided by looking at fixed
scalars which {\it do not} obey the minimally coupled Klein-Gordon
equation. An example of this is the size of the
$T^4$. To analyze this one has to substitute
\eqn\ceight{G_{IJ}^S = e^{2\nu} \delta_{IJ}}
in equation \ctwo. 
Since this is the trace part of the ten dimensional graviton, it
mixes with the dilaton. The requirement that no {\it five 
dimensional} dilaton is emitted means that one has to set
$ \phi = 2 \nu$ \callanfixed. It is clear from \ctwo\ that $\nu$
couples only to the quartic term in the action. 
The DBI action becomes, upto terms linear in $\nu$ ,
\eqn\cnine{S = {\Te \over 2}\int dx^0 dx^5 [ 1 + 
\half \partial_+ X^I \partial_-
X_I - {1 \over 8} (\partial_+ X)^2 (\partial_- X)^2
- {1 \over 4} \nu (\partial_+ X)^2 (\partial_- X)^2]}
Thus the emission of a quanta of $\nu$ has the important
characteristic that the lowest order process involves four open
string fields. For the same reason the effective string tension
$\Te$ does not scale out and the answer depends on $\Te$.

Once again, the field $\nu$ has to be treated as a function of $x^0$
and $x^5$ only. It was shown in \callanfixed\ that the emission rate
is in exact agreement with the semiclassical answer provided $\Te = 1/
(2 \pi \ap Q_5)$ and $L_{eff} = 2\pi Q_1 Q_5 R$ as expected from
earlier considerations.

\subsec{Absorption by 3-branes}

In the above examples, there is as yet no complete derivation of the 
effective string model from the microscopic model of branes. From this
point of view it is useful to consider parallel branes of a single
type which lead to nonsingular geometries. An example is the self dual
3-brane in Type IIB theory.  Consider for example 
the absorption of dilatons by an extremal 3-brane.
The relevant interaction comes from a term involving two
worldvolume gauge fields going into a dilaton \igorc\ 
\eqn\fone{S_{int} \sim -{\kappa \over {\sqrt{2}}}\int d^4 x~\phi
(x^\alpha
, X^i){\rm Tr}[\partial_\alpha A_\beta \partial^\alpha A^\beta]}
where $x^\alpha$ are the four brane coordinates and $X^i$ denote the
transverse coordinates. $A_\alpha$ is the gauge field.

The emission of dilatons in some given partial wave $l$ is described
by the following term in the Taylor expansion of $\phi(x^\alpha, X^i)$
in powers of $X^i$ \igorc,\igore\
\eqn\fonea{ L_l \sim (\partial_{i_1}\cdots \partial_{i_l} \phi)_0
~{\rm Tr}[<X^{i_1} \cdots X^{i_l}> F_{\alpha \beta}^2]}
where $<X^{i_1}\cdots X^{i_l}>$ denotes the projection of the products
to the definite angular momenta, and the derivatives on $\phi$ are
evaluated at the location of the brane at $X^i = 0$. Similar
considerations apply to other closed string states like the
longitudinally polarized graviton and the RR scalar \igore.

It is clear from the classical solution that the expansion parameter
in the classical absorption cross-section of a wave of frequency
$\omega$ is the combination $\omega (\ap^2 gN)^{1/4}$. It has been
shown in \igore\ that for the $l$th partial wave the answer is
\eqn\foneb{ \sigma^{l}(\omega) \sim \omega^{(4l + 3)} (\ap^2 gN)^{l +
2}}
Remarkably, for $l = 0,1$ the cross-section exactly matches the
D-brane answer, whereas for $l = 2,3$ the two answers differ by a
numerical factor \igore.

\newsec{Strategy for higher loops}

In the following we will examine lowest order open string corrections
to absorption/emission processes from slightly non-extremal single
branes and parallel branes. In these situations explicit string theory
calculations can be performed and these are directly relevant to the
absorption by three branes. Moreover the 1-brane 5-brane system
reduces to a single long D-string at low energies. Consequently some
of our results will be relevant for this case as well.

To appreciate the issue of loop corrections, consider for simplicity
black holes which contain a single length scale in the
problem. Examples are extremal branes (the 3-brane in particular) and
the five dimensional black hole in the special case $r_1 = r_5 = r_N =
R$. Let us call this length scale $l$. In the classical solution the
string coupling can enter only through this length scale $l$ which is
typically given by the form
\eqn\addone{ l^{(d-3)} \sim g Q \ap^{(d-3/2)}}
where $d$ denotes the number of non-compact dimensions. It is then
clear that the classical absorption cross-section has to be of the
form
\eqn\addtwo{\sigma_{class} \sim l^{d-2} F(\omega^{(d-3)}g Q \ap^{(d-3/2)})}
On general grounds we expect that this classical answer should agree
with the D-brane answer when $gQ$ is large. However the above
expression shows that for sufficiently small $\omega$ one may have the
factor $\omega^{(d-3)}g Q \ap^{(d-3/2)}$ small even if $gQ$ is large
so that one may imagine performing a Taylor expansion of the function
$F$, which then becomes a power series expansion in the string
coupling $g$ as well \igorc. 
The spectacular success of the tree level D-brane calculations of the
absorption cross-section then means that the {\it lowest order} term
in this expansion has been shown to agree with the {\it lowest order}
term in D-brane open string perturbation theory.
 
The puzzle regarding this agreement of D-brane and classical
calculations may be now restated as
follows : In the classical limit a higher power of the string coupling
comes with a higher power of the energy in a specific way dictated 
by \addtwo. On the other hand, on the D-brane side these higher powers
of coupling are to be obtained in open string perturbation theory
and there is no {\it a priori} reason why this should also involve
higher powers of energy in precisely the same way.

This implies that the following two kinds of loop corrections must be
absent for the correspondence to work. (1) For a given absorption
or emission process open string loop
diagrams with the same external states must be suppressed at low
energies compared to the tree diagram. (2) Suppose we concentrate on
emission of some given closed string state and let the leading order
tree process give a cross-section $\sigma \sim g^\alpha$ with some
energy dependence. For $\alpha$ large enough it is possible that there
is a string loop process with the same dependence on the coupling
\foot{The importance of this was emphasized to me by I. Klebanov.}. 
This must, of course, involve external states which are {\it
different} from the tree process. For processes where the D-brane
tree level calculation gave the correct answer such loop corrections
must be suppressed at low energies.

What we want to calculate is the (open) string loop correction to the
effective action of the massless open string modes in the presence of
a background closed string which we will take to be the ten
dimensional metric or the dilaton.  As we will see soon these diagrams
are generically nonzero. However it must be remembered that we have
compared processes only at low energies.  We have to see whether there
are loop corrections at such low energies.

For S-wave emission/absorption low energies meant the minimal powers
of momenta of the open string states and no powers of momenta for the
closed string states. In oher words when the effective action is
expressed in the presence of closed string fields {\it on the brane}
there were no derivatives with respect to either the longitudinal or
transverse directions.  To examine whether these specific processes
receive a one-loop correction it is sufficient to consider the
effective action in {\it flat} space.  Then the curved space effective
action may be obtained by simply replacing the flat metric by the
curved metric. This follows from general coordinate invariance - in
fact the equivalence principle.  For example for the case of a
D-string, if we know the term in the flat space effective action
\eqn\done{
\partial_+ X^I \partial_- X^J \delta_{IJ}}
Then in the curved space effective action, $\delta_{IJ}$ may be
replaced by any tensor made out of the metric, e.g.
\eqn\dtwo{
\partial_+ X^I \partial_- X^J [G_{IJ} + R_{IJ} + \cdots]}
where $R_{IJ}$ is the Ricci tensor. However only the first term 
in \dtwo\ does
not involve any derivative of the metric. Thus at low energies only
the first term is relevant, which may be obtained by simply replacing
the flat metric $\delta_{IJ}$ by the curved metric $G_{IJ}$. In a
similar fashion once we figure out the term in the flat space
effective action 
\eqn\dthree{
\partial_+ X^I \partial_+ X^J \partial_-X^K \partial_- X^L \delta_{IJ}
\delta_{KL}}
we will know the loop corrections to the term which was responsible
for emission of fixed scalars. 
The fact that it is sufficient to consider the effective action in
flat space implies that there is enhanced supersymmetry in the 
problem \foot{This was pointed out by S. Shenker.}.

We are interested in low energies of the open string
modes as well. This means that in a term with some number of open string
fields we keep the term with the minimal number of derivatives on
these fields. This means we are not interested in terms like 
\eqn\dfour{
\partial^2 X^I \partial^2 X^J~~~~~~~~\partial^2 X^I \partial^2 X^J
\partial_+ X^K \partial_- X^L \cdots}
and so on.

To find whether higher loop processes involving different external
states may correct a tree level emission/absorption of some given
closed string state, we have to include insertions of closed string
backgrounds. Nevertheless the calculation will be simplified by
requiring that we look at the term which has the required powers of
the momentum.

Loop diagrams with insertions of closed string states are also
essential to study possible corrections to the absorption/emission
in higher partial waves. This is because for such processes the tree
level term itself involves derivatives on the closed string fields
(see e.g. \fonea).

There are, however, other S-wave processes where the low energy curved
space effective action cannot be read off from the flat space action.
These involve situations where the worldsheet is curved. Recall that
we are using the static gauge, so that certain components of the 
background metric become worldsheet metrics. In the five dimensional
black hole an example is the component $h_{55}$. It follows from the
Born-Infeld action that this has couplings of the form
\eqn\addthree{ h_{55} [(\partial_+ X)^I (\partial_+X_I)
+ (\partial_- X)^I (\partial_- X_I)]}
i.e. this couples to a sum of {\it chiral} operators. Such operators
cannot appear in the DBI action in flat backgrounds. For such fields
the loop corrections cannot be obtained from the flat space terms.

\newsec{The One Loop Calculation : Flat space}

We will first analyze open string loop diagrams without any closed
string insertion. We will use the light cone gauge Green-Schwarz
formalism. To do this, however, we have to consider the brane
directions to be euclidean and one of the transverse directions to be
time. Such a framework has been discussed in
\ref\GREEN{M. Green and M. Gutperle, Phys. Lett. B377 (1996) 28,
hep-th/9602077; Nucl. Phys. B476 (1996) 9604091; K. Hamada,
hep-th/9612234}.
For a $p$-brane we thus have $(X^1 \cdots X^{p+1})$ to be
the longitudinal directions, $(X^{p+2} \cdots X^8, X^\pm = X^0 \pm X^9)$
as the transverse directions. This specifies the boundary conditions
for the bosonic fields of the open string in the standard fashion.
The boundary conditions for the fermions are determined by the
requirement that half of the supersymmetries are unbroken and are given
by
\eqn\aone{
S_+^a (\sigma,\tau) = M^a_b S_-^b (\sigma.\tau)~~~~~~~~~~~~~
\sigma = 0,\pi}
The matrix $M$ may be written in terms of the ten dimensional gamma
matrices \GREEN,
\ref\grav{ S. Gubser, A. Hashimoto, I. Klabanov and J. Maldacena,
Nucl. Phys. B472 (1996) 231, hep-th/9601057.},
\ref\garmyers{ M. Garousi and R. Myers, Nucl. Phys. B475 (1996) 193,
hep-th/9603194.}.
For our purposes we will need the properties
\eqn\atwo{\eqalign{
& M^T \gamma^I M =  - \gamma^I~~~~~~~~~(I = (p+2) \cdots 8) \cr
& M^T \gamma^\alpha M  =  \gamma^\alpha~~~~~~~~(\alpha = 1 \cdots
p)\cr
& M^T M = I}}
The vertex operators for the open string massless states with polarizations
in the Dirichlet and Neumann directions are given by (respectively)
\eqn\athree{\eqalign{
& V_D = \zeta_I (k)
 (\partial_\sigma X^I - S_+ \gamma^{I\alpha} S_+ k_\alpha)e^{ikX}\cr
& V_N = \zeta_\beta (k)(\partial_\tau X^\beta - S_+ \gamma^{\beta\alpha} 
S_+ k_\alpha)e^{ikX}}}
where we have used the fact that the momenta of the open string states
are always in the Neumann direction.

\subsec{The quadratic term}

Let us first consider the terms in the one (open string) loop
contribution to the effective action which involve two open string fields.
When both the open string operators are on the same boundary we have
a term like
\eqn\afour{
K_{(2,0)}{\rm Tr}[ V (k_1) \Delta V (-k_1) \Delta]}
where $\Delta$ denotes the open string propagator and $V$ stands for
either $V_D$ or $V_N$.
Here $K_{(2,0)}$ is the Chan-Paton factor. For a single brane this is just
the $U(1)$ charge, whereas for multiple branes we have the usual trace
over the $U(N)$ group generators.

Since we are dealing with an oriented string theory one has to add the
contribution from the case where they attach to the two different
boundaries, which is obtained by insertion of two twist operators
$\Omega$
\eqn\afive{K_{(1,1)}{\rm Tr}[V (k_1) \Omega \Delta V (-k_1) \Omega \Delta]}
The Chan Paton factor $K_{(1,1)}$ is nonzero only for the overall
$U(1)$.

It is well known that the terms \afour\ and \afive\ individually
vanish. This happens because the trace involves a trace over the zero
modes of the fermion fields $S_0^a$. Only products of at least eight
zero modes have a nonzero trace, while the terms like \afour\ can
involve only four such zero modes. This is in fact a reflection of
supersymmetry of the underlying theory.

Thus, to lowest order in the energy there are no one loop correction
to the effective action involving two open string fields. This result
is valid for parallel branes of any type. In particular, this
establishes that the lowest order S-wave dilaton absorption by
3-branes is not renormalized to this order. In so far as the 1-brane
5-brane system can be regarded as a single long D-string, this also
implies that the emission/absorption of minimal scalars in the five
dimensional black hole is not renormalized.

This result is an explicit verfication of the non-renormalization
arguments in \Maldacenab.
 
\subsec{The quartic term}

Now consider terms in the effective action which involve four open
string fields - again in flat space. The planar annulus amplitude
is given by
\eqn\six{
K_{(4,0)}
{\rm Tr}[ V (k_1) \Delta V (k_2) \Delta V (k_3) 
\Delta V (k_4) \Delta]}
where $K_{(4,0)}$ is the relevant Chan-Paton factor.
Since each vertex operator involves two fermion zero modes, \six\
has a single term with eight fermion zero modes so that the vertex
operators appearing  in \six\ may be replaced by
\eqn\seven{
V \sim S_0 \gamma^{\mu\alpha} S_0 k_\alpha \zeta_\mu (k) e^{ikX}}
We are looking for a term which has four powers of open string momenta -
one for each open string field. This means that we can ignore the
$e^{ikX}$ factor in \seven\ and use
\eqn\eight{
V \sim S_0 \gamma^{\mu\alpha} S_0 k_\alpha \zeta_\mu (k)}
The final nonzero result can be easily computed following the procedures in
for example 
\ref\GSW{  M. Green, J. Schwarz and E. Witten, {\it Superstring Theory,
Vol. 2} (Cambridge University Press, 1986)}.
Let us call this ${\cal A}_2$.

As before, we have to add the contributions from the non-planar diagrams
which may be obtained by putting in an even number of twist operators in
\six. For example a nonplanar diagram with three vertices on
one of the boundaries and one vertex on the other boundary is given by
\eqn\aeight{K_{(3,1)}{\rm Tr}[ V (k_1)
\Omega \Delta V (k_2) \Omega \Delta V (k_3) \Delta V (k_4) \Delta]}
where $K_{(3,1)}$ is the relevant Chan-Paton factor in this case.
Using $\Omega^2 = 1$ the effect of a twist operator is easily seen to
result in a change of sign of 
of the odd oscillators, $\alpha_n \rightarrow (-1)^n \alpha_{n},
S_n \rightarrow (-1)^n S_n$ in all the vertex operators on one of the
boundaries. For \aeight\ this means that the odd oscillators
in $V (k_2)$ have to be flipped.

However, as we have just found, at low open string energies there are
no oscillators in the vertex operators ! Thus,
at low energies,  the individual contributions
of each of the nonplanar diagrams is the same as that of the planar
diagrams apart from (i) combinatoric factors and (ii) Chan-Paton factors.
There are $2$ terms with all the vertices on the same boundaries, $8$
terms with three vertices on one boundary and one vertex on the other
boundary and $6$ terms with two vertex operators on each boundary.
Thus we get the net contribution
\eqn\nine{(2K_{(4,0)} + 8 K_{(3,1)} + 6 K_{(2,2)}) {\cal A}_2}
This is in general nonzero.

However, when we have a single brane, the gauge group is $U(1)$ and
the two ends of the open string going around the loop have equal and
opposite charges $\epsilon$. In this case 
\eqn\ninea{ K_{(4,0)} = - K_{(3,1)} = K_{(2,2)} = \epsilon^4}
and the net contribution \nine\ vanishes.

We emphasize that this result is true for the lowest order term in the
open string energy. At higher energies the $e^{ikX}$ factors in $V$
cannot be set to $1$ and the vertex operators will contain
oscillators. As a result the magnitude of a non-planar diagram will
be different from a planar diagram and the above cancellation will
not take place.

Nonrenormalization properties of $F^4$ terms have been shown for the
Type I superstring and follow from cancellations between the annulus
and the mobius strip diagrams \ref\tseytlin{A. Tseytlin ,
Phys. Lett. B367 (1996) 84, hep-th/9510173; Nucl. Phys. B467 (1996)
383, hep-th/9512081.}. Similar nonrenormalization theorems are also
required for the consistency of M(atrix) theory \ref\BFSS{T. Banks,
W. Fischler, S. Shenker and L. Susskind, hep-th/9610043.}. However, as
we discuss below, the terms which are required to be protected in this
case are of a rather different nature than the terms we have been
discussing.

\subsec{Relationship to brane dynamics}

For single branes, the one loop
calculation in the previous section is closely related to the
dynamics of {\it two} branes which are moving and have waves on it as well.
In this situation, the force between these branes may be calculated
by evaluating an annulus diagram in the open string theory where the
two boundaries of the worldsheet are attached to the two {\it different}
branes. The worldsheet theory of the open strings connecting these
branes now contain the terms
\eqn\eone{
e_1 \int_{\sigma = 0} d\tau A_I (X^\alpha) (\partial_\sigma X^I + \cdots)-
e_2 \int_{\sigma = \pi} d\tau A_I (X^\alpha) (\partial_\sigma X^I + \cdots)}
where $\alpha$ denotes a Neumann direction and $I$ denotes a Dirichlet
direction and the dots represent fermionic terms.
It is then easily seen that the annulus amplitudes we evaluated in
the previous section are obtained by expanding the partition function
in the presence of the additional terms \eone\ in powers
of the backgrounds $A_I(x^\alpha)$ and (i) setting $e_1 = e_2$ and
(ii) setting the distance between the branes to zero.

Since we were interested in the situation where $\partial_\alpha A_I$
are constant, the $A_I$'s we need to consider are those which are
at most linear in the $X^\alpha$'s.

When the background field is only a function of the time coordinate,
i.e. $A_I(x^0)$ we have the two branes moving with velocities $e_1$
and $e_2$, a situation analyzed in 
\ref\bachas{C. Bachas, Phys. Lett. B374 (1996) 37,hep-th/9511043}.
In this case the
zero result we found to order $A_I^4$ is simply a special case of the
general result that the force between BPS branes vanish when the
relative velocity is zero - which follows from Lorentz invariance and
the fact that static BPS branes have no forces between them regardless
of the distance of separation.

Our result is, however, more general than this situation. For a
general $p$-brane we can transform our problem to the case of
uniformly moving branes in only some special situations. Consider for
example the case of $p = 1$. In that case the general form of the 
background is
\eqn\etwo{
A_I(x^0,x^1) = E_I x^0 + B_I x^1}
When $E_I$ and $B_I$ are parallel to each other so that 
$E_I = E {\hat e}_I$ and $B_I = B {\hat e}_I$ the term in the action
is simply
\eqn\ethree{
(EX^0 + BX^1)\partial_\sigma ({\hat e}_I X^I)}
Now we can perform a Lorentz transformation in the coordinates 
(provided $E^2 > B^2$) to go to a new time variable $Y^0$ in terms of
which we simply have
\eqn\efour{
({\sqrt{E^2 - B^2}}) Y^0 \partial_\sigma ({\hat e}_I X^I)}
This is exactly the case which may be mapped to uniformly moving
branes.

When $E_I$ and $B_I$ are not parallel this cannot be done any more and
we have a situation which is physically quite different. There are no
obvious reasons why the one loop corrections to the effective action
vanish.  It is, however, tempting to speculate that terms in the
effective action with higher powers of the open string fields also do
not receive any loop corrections. This would be relevant to the
emission of particles with angular momentum 
\ref\malstromb{J. Maldacena and A. Strominger, hep-th/9702015.}.

The above discussion shows that the nonrenormalization property we are
looking for is of a different nature than that required in M(atrix)
theory. In the latter context one is interested in the case where
two branes are moving with respect to one another. There is no
tree level term and the entire contrbution is expetced to arise from
one loop. In our case there is a tree level term which needs to be
protected from loop corrections.

It may appear that our vanishing result for the abelian situation is
simply a reflection of the fact that there are no charged fields in
the theory. However we are looking at higher energy processes where
the nonzero dipole moment of the open string might have led to
nontrivial effects. In fact, by T-duality the problem may be related
to the problem of open strings in constant electromagnetic backgrounds
\foot{In doing this it must be remembered that the open string momenta
continue to be only in the Neumann directions.}. In this T-dual
situation it is well known that there is a one loop correction to the
neutral {\it bosonic} string partition function in the presence of
abelian constant fields. In fact one has
\ref\callannappi{A. Abouelsaood, C.G. Callan, C.R. Nappi and
S.A. Yost, Nucl. Phys. B280 [FS18] (1987) 599; C. Bachas and
M. Porrati, Phys. Lett. B296 (1992) 77.}
\eqn\aten{ Z_{1-loop}(F) = {\rm det}[1 + F]~Z_{1-loop} (0)}
$Z_{1-loop}(0)$, the zero field partition function is nonzero (in fact
infinite due to the tachyon) for the bosonic string. For the
superstring a similar formula is likely to hold \foot{I would like to
thank C. Callan for a discussion on this point.}. Then $Z_{1-loop}(F)$
would vanish because $Z_{1-loop}(0)$ vanishes as a result of
supersymmetry. This seems to suggest that in our problem the higher
order terms in the open string fields would also vanish.

\newsec{Loop effects with closed string insertions}

We now consider the effect of closed string insertions on loops which
could give contributions to the absorption processes discussed
above. Potentially a one loop diagram with lower number of open string
fields can contribute, in the same order of the string coupling $g$, 
to a process where the tree level involves a higher number of open
strings. 

It is clear that there cannot be any such correction to the emission
of S-wave minimally coupled scalars in the five dimensional black
hole. The tree level process for this is of order $g$ while the loop
level process starts at order $g^2$

Consider for example the annulus diagram with two open strings and one
closed string. The vertex operator for a massless bosonic closed string
state is given by
\eqn\htwo{(\partial_+ X^\mu - S_+ \gamma^{\mu\alpha}S_+ k_\alpha)
(\partial_- X^\nu - \eta S_+ \gamma^{\nu\beta}S_+ k_\beta)~e^{ikX}}
where $\eta = +1$ for Neumann directions and $\eta = -1$ for Dirichlet
directions.  Since the closed string vertex has four fermionic fields,
the term with the minimal power of momenta is obtained by replacing
the vertex operators by their fermionic terms only and furthermore by
replacingthe fermionic fields by their zero modes. Schematically this
would give rise to a term in the effective action of the form
\eqn\hone{g^2 \partial X \partial X \partial\partial \phi}
where $\phi$ denotes a massless closed string field and $X$ an open
string field. At one loop this can contribute to the absorption of
fixed scalars in the 1-brane 5-brane system and absorption of $l = 2$
modes in the 3-brane system, since both these processes go as $g^2$ at
the tree level. Note that at one loop there is no overall dilaton
factor in front of the effective action as a result of which fixed
scalars can interact with just two open string fields.

When we have single branes (so that the gauge group is abelian), the
planar diagram cancels the nonplanar diagram for the reason discussed
above : the vertex operators do not involve any oscillators so that
the insertion of a twist operator does not change the magnitude of the
diagram. In so far as the effective string model is trustworthy at
this level, this would mean that this does not alter the
absorption/emission of fixed scalars. Similarly, this shows that the
$l = 2$ absorption by a {\it single} 3-brane does not receive a
$O(g^2)$ contribution from one loop.

This argument does not {\it a priori} apply to the case of $N$
3-branes because of non-trivial Chan-Paton factors.
One may wonder whether a nonzero contribution from such
one loop diagrams may account for the difference between the brane and
classical results in this case \igore. However this cannot be the
case, since the absorption cross-section obtained after summing over
the open string states would go as $ g^4 \ap^6 \omega^7$ as opposed to
the tree level result $g^4 \ap^8 \omega^{11}$. This is
because the tree level diagram has more open string states which give
higher powers of momentum from the phase space factors. This would
make the one loop result dominate over the tree level result  at low
energies. While we do not have a good reason why this particular one
loop contribution vanishes even for $N$ 3-branes, we believe that the
answer has got to do with the fact that to this order the 3-brane
system behaves as a collection of $N^2$ abelian fields.

Finally we consider higher order (in string loop) corrections to terms
in the 3-brane effective action which are responsible for the higher
partial wave absorptions. The term for the $l$th partial wave is given
in \fonea\ and involves $l$ derivatives on the closed string field and
one derivative each on two fo the $(l+2)$ open string fields. We want
to see whether there could be one-loop corrections to such terms.

Our analysis shows that any one-loop contribution must have at least
four powers of the momentum, which may be distributed among the open
and closed string fields, since there must be at least eight fermion
zero modes. This immediately shows that there cannot be one-loop
contributions to terms like \fonea\ for $l = 0, 1$ since at tree level
they do not have the necessary powers of momenta. Any loop correction
to such processes would be therefore suppressed at low energies. On
the other hand there could be in general nonzero contributions to $l
\geq 2$ partial waves. Furthermore for such terms there is no
cancellation between planar and nonplanar diagrams even for single
branes since there are at least two open string fields with no momenta
accompanying them. This could arise when the $\partial_\sigma X$ term
in the corresponding vertex operators contribute to the one loop
diagram as the necessary fermionic zero modes are already supplied by
two open and one closed string field. This term necessarily involves
oscillators which are indeed affected by the action of a twist
operator. Detailed analysis of such diagrams are need : perhaps this
is the reason why there is a discrepancy for the $ l \geq 2$
absorption cross-sections.

\newsec{Remarks}

Some of the above results were derived for single or parallel branes
and are directly applicable to the 3-brane system. They suggest that
for this system some of the absorption processes do not receive
corrections from one loop. This could be the reason why the results
agree with the semiclassical answers.

The five dimensional (or the four dimensional) black hole is described
in terms of brane systems for which it is difficult to do microscopic
calculations. However we believe that some of the results we have
presented for single branes are relevant to this case as well since
most of the results in fact follow from an effective string
picture. Assuming that the system can be indeed modelled by such a
long D-string, our results give evidence for the nonrenormalization of
emission/absorption of S-wave minimally coupled scalars and fixed
scalars. However the validity of this effective string picture at the
loop level need to be examined.

Finally we would like to emphasize that even if one is able to prove
that similar nonrenormalization theorems protect the lowest order tree
level results, it does not immediately follow that the information
loss problem is solved. What is really missing here is a clear
understanding of the horizon physics in the problem. The curvature of
the background comes from summing over open string loops. It remains
to be seen whether there are {\it non-perturbative} effects which
spoil correspondence of this perturbation theory with the physics of
the horizon. In view of the spectacular agreement of the results it is
tempting to speculate that there are none.

\newsec{Acknowledgements} A very preliminary version of this work was
presented at the ``Workshop on String Theory, Field Theory and Quantum
Gravity'' held at Puri in December 1996. I thank the local organizers
for organizing a wonderful conference.  I would like to thank
A. Dabholkar, A. Jevicki, J. Maldacena, A. Sen, S. Shenker and
L. Susskind for valuable discussions and A. Tseytlin for a
correspondence.  I especially thank Samir Mathur for many discussions
and I. Klebanov for enlightening discussions and comments on the
manuscript. I would also like to thank the Physics Departments of
Brown University and Rutgers University for hospitality during the
completion of this work.

\listrefs
\end